# Possible dynamics of the Tsallis distribution from a Fokker-Planck equation (I)

Du Jiulin[*]

*Department of Physics, School of Science, Tianjin University, Tianjin 300072, China*

**Abstract**

The dynamical property of the Tsallis distribution is studied from a Fokker-Planck equation. For the Langevin dynamical system with an *arbitrary* potential function, Markovian friction and Gaussian white noise, we show that no possible nonequilibrium dynamics can use the Tsallis distribution for the statistical description. The current form of the Tsallis distribution stands for a simple isothermal situation with no friction and no noise.

**PACS** numbers: 05.20.-y, 05.10.Gg

Nonextensive statistical mechanics since it started in 1988 has obtained very wide applications in many interesting scientific fields. In principle, almost all the formulae and the theory using Boltzmann-Gibbs statistics so far could be generalized under this framework [1-4] (for more details, see http://tsallis.cat.cbpf.br/biblio.htm ). However, the questions, such as under what circumstances, e.g. which class of nonextensive systems and under what physical situation, should the nonextensive statistical mechanics be used for their statistical description have been long-standing [5-13]. The purpose of this work is to investigate the question whether there is a possible dynamics that can be described by the basic form of Tsallis distribution, $f \sim [1-(1-q)\beta H]^{\frac{1}{1-q}}$, in nonextensive statistical mechanics.

Tsallis proposed the entropy in 1988 as a generalization of the Boltzmann-Gibbs

---
[*] E-mail address: jiulindu@yahoo.com.cn

entropy [14], given by

$$S_q = -k\sum_i p_i^q \ln_q p_i ,\qquad(1)$$

where $k$ is Boltzmann constant, the set $\{p_i\}$ are the probabilities of the microscopic configurations of the system under investigation, the parameter $q$ is real number, the $q$-logarithm is defined as

$$\ln_q x \equiv \frac{x^{1-q}-1}{1-q} \quad (x>0; \ln_1 = \ln x),\qquad(2)$$

the inverse function, the $q$-exponential, is

$$\exp_q x \equiv [1+(1-q)x]^{\frac{1}{1-q}} \quad (\exp_1 x = \exp x),\qquad(3)$$

if $1+(1-q)x >0$ and by $\exp_q x = 0$ otherwise. The entropy $S_q$ ($q \neq 1$) is *nonextensive*; namely, if a system composed by two probabilistically independent parts $A$ and $B$, i.e., $p_{ij}^{A\oplus B} = p_i^A p_j^B \ \forall(ij)$, then the Tsallis entropy of the system is

$$S_q(A\oplus B) = S_q(A) + S_q(B) + (1-q)k^{-1}S_q(A)S_q(B).\qquad(4)$$

Under this framework, one leads to the basic form of the Tsallis distribution, $f\sim [1-(1-q)\beta H]^{\frac{1}{1-q}}$, with the Lagrange parameter, $\beta = 1/kT$, and the Hamiltonian $H$, in the nonextensive statistical mechanics. Clearly, only if taking $q=1$, does the Tsallis distribution become Boltzmann distribution, $f\sim \exp(-\beta H)$.

We begin with a quite general Langevin equations for the dynamics of a set of variables $\mathbf{a}=\{a_i\}$. The abbreviated form is

$$\frac{d\mathbf{a}}{dt} = \mathbf{V}(\mathbf{a}) + \mathbf{F}(t),\qquad(5)$$

where $\mathbf{V}(\mathbf{a})=\{V_i(\{a_j\})\}$ is a set of some given function of the variables $\mathbf{a}$. One may require that the noise-free part $V_i$ of the dynamics is Markovian and that the noise $\mathbf{F}(t)=\{F_i\}$ is a set of white and has a Gaussian distribution, with zero mean and the delta-correlated second moment matrices,

$$\langle \mathbf{F}(t)\mathbf{F}(t')\rangle = 2\mathbf{B}\delta(t-t');\ \langle \mathbf{F}(t)\rangle = 0,\qquad(6)$$

where the quantity is $\mathbf{B}=\{B_{ij}\}$; $i,j=1,2,\ldots n$.

The question we want to study here is whether or not there is a dynamics of the Langevin equation Eq.(5) and Eq.(6) that can be described by the Tsallis distribution $f \sim [1-(1-q)\beta H]^{\frac{1}{1-q}}$. Rather than looking for a general solution of these equations, we usually ask for the probability distribution $f(\mathbf{a}, t)$ of the values of $\mathbf{a}$ at time $t$. What we want is the average of this probability distribution over the noise. The Fokker-Planck equation, corresponding to the dynamics governed by the above Langevin equations [15], for the noise-averaged distribution function $f(\mathbf{a},t)$ can be written by

$$\frac{\partial}{\partial t} f(\mathbf{a},t) = -\frac{\partial}{\partial \mathbf{a}} \cdot \mathbf{V}(\mathbf{a}) f(\mathbf{a},t) + \frac{\partial}{\partial \mathbf{a}} \cdot \mathbf{B} \cdot \frac{\partial}{\partial \mathbf{a}} f(\mathbf{a},t). \qquad (7)$$

The first term on the right-hand side is what one has on the absence of noise; the second term accounts for the averaged effects of the noise. At this point, $\mathbf{B}$ can be any function of $\mathbf{a}$.

So far nothing has been said about requiring that $f(\mathbf{a},t)$ must approach an equilibrium distribution at long times. In fact, not much is known in general about the long times stationary-state solution of an *arbitrary* Fokker-Planck equation. But we can usually do guess at a stationary-state solution, put it into the equation, and then search for if the guess is compatible with $\mathbf{V}(\mathbf{a})$ and $\mathbf{B}$. If it is so, then the stationary-state solution can describe the long-times dynamics govern by the above Langevin equation. Our question therefore becomes to find whether there is the possible $\mathbf{V}(\mathbf{a})$ and $\mathbf{B}$ that have the full physical property compatible with the Tsallis distribution.

One assumes the Tsallis distribution function $f \sim [1-(1-q)\beta H]^{\frac{1}{1-q}}$ to satisfy the stationary-state Fokker-Planck equation,

$$-\frac{\partial}{\partial \mathbf{a}} \cdot \mathbf{V}(\mathbf{a}) f(\mathbf{a}) + \frac{\partial}{\partial \mathbf{a}} \cdot \mathbf{B} \cdot \frac{\partial}{\partial \mathbf{a}} f(\mathbf{a}). \qquad (8)$$

Equivalently, it can be written as the following partial differential equation,

$$-\sum_{i=1}^{n} \frac{\partial}{\partial a_i}(V_i f) + \sum_{i,j=1}^{n} \frac{\partial}{\partial a_i}\left(B_{ij} \frac{\partial f}{\partial a_j}\right) = 0, \qquad (9)$$

or, furthermore, it is

$$-\sum_{i=1}^{n}\frac{\partial}{\partial a_i}(V_i f)+\sum_{i,j=1}^{n}\left[B_{ij}\frac{\partial^2 f}{\partial a_i \partial a_j}+\left(\frac{\partial B_{ij}}{\partial a_i}\right)\left(\frac{\partial f}{\partial a_j}\right)\right]=0 \tag{10}$$

The Tsallis distribution function as given in nonextensiv statistical mechanics so far is

$$f \sim [1-(1-q)\beta H]^{\frac{1}{1-q}} \equiv R^{\frac{1}{1-q}}, \tag{11}$$

where the Hamiltonian is $H = H(\mathbf{a})$ and the Lagrange parameter $\beta = \beta(\mathbf{a})$. Put it into Eq.(10), the equation becomes

$$-f\sum_{i=1}^{n}\frac{\partial V_i}{\partial a_i}+f^{-1}\sum_{i=1}^{n}V_i\frac{\partial(\beta H)}{\partial a_i}-f^{-1}\sum_{i,j=1}^{n}\frac{\partial(\beta H)}{\partial a_i}\frac{\partial B_{ij}}{\partial a_i}-\sum_{i,j=1}^{n}B_{ij}\frac{\partial}{\partial a_i}\left[f^{-1}\frac{\partial(\beta H)}{\partial a_j}\right]=0, \tag{12}$$

or, it can be written in terms of $R$ as

$$-R\sum_{i=1}^{n}\frac{\partial V_i}{\partial a_i}+\sum_{i=1}^{n}V_i\frac{\partial(\beta H)}{\partial a_i}-\sum_{i,j=1}^{n}\frac{\partial B_{ij}}{\partial a_i}\frac{\partial(\beta H)}{\partial a_i}-\sum_{i,j=1}^{n}B_{ij}\frac{\partial^2(\beta H)}{\partial a_i \partial a_j}$$

$$+qR^{-1}\sum_{i,j=1}^{n}\frac{\partial(\beta H)}{\partial a_i}B_{ij}\frac{\partial(\beta H)}{\partial a_j}=0. \tag{13}$$

If this equation can determine a relation with full physical property between $\mathbf{V}(\mathbf{a})$ and $\mathbf{B}(\mathbf{a})$, which may be called a generalized fluctuation-dissipation theorem, then the dynamics of the Langevin equations that can be described by the Tsallis distribution is found.

We consider the dynamics of the two-variable Brownian motion of a particle, with mass $m$, in an *arbitrary* potential $\varphi(x)$ (whether long-range or short-range force); the Langevin equations about the coordinate $x$ and the momentum $p$ are

$$\frac{dx}{dt}=\frac{p}{m}, \quad \frac{dp}{dt}=-\frac{d\varphi}{dx}-\zeta\frac{p}{m}+F_p(t), \tag{14}$$

where $\zeta$ is the frictional coefficient. The noise is Gaussian and it is delta-function correlated,

$$\langle F_p(t)F_p(t')\rangle = 2B\delta(t-t'). \tag{15}$$

The quantities that go into the Fokker-Planck equation are

$$\mathbf{a}=\begin{bmatrix}x\\p\end{bmatrix}, \quad \mathbf{V}(\mathbf{a})=\begin{bmatrix}p/m\\-d\varphi/dx-\zeta\,p/m\end{bmatrix},$$

$$\mathbf{F}(t)=\begin{bmatrix} 0 \\ F_p(t) \end{bmatrix}, \quad \mathbf{B}=\begin{bmatrix} 0 & 0 \\ 0 & B \end{bmatrix}. \tag{16}$$

Then the Fokker-Planck equation (7) becomes

$$\frac{\partial f}{\partial t} = -\frac{\partial}{\partial x}\frac{p}{m}f + \frac{\partial}{\partial p}\left(\frac{d\varphi}{dx} + \zeta\frac{p}{m}\right)f + B\frac{\partial^2 f}{\partial p^2}, \tag{17}$$

and the stationary-state equation (13) for the Tsallis distribution becomes

$$\frac{p}{m}\left(\frac{d(\beta\varphi)}{dx} + \frac{p^2}{2m}\frac{d\beta}{dx}\right) - \frac{\beta p}{m}\left(\frac{d\varphi}{dx} + \frac{\zeta p}{m}\right) + \frac{\zeta}{m}R - \frac{\beta B}{m}\left(1 - q\frac{\beta p^2}{m}R^{-1}\right) = 0. \tag{18}$$

or, it can be written as

$$\frac{\zeta}{m}R^2 + \left(\frac{1}{2m^2}\frac{d\beta}{dx}p^3 - \frac{\zeta\beta}{m^2}p^2 + \frac{\varphi}{m}\frac{d\beta}{dx}p - \frac{B\beta}{m}\right)R + q\frac{B\beta^2}{m^2}p^2 = 0. \tag{19}$$

Substituting into $R = 1 - (1-q)\beta H$, where the Hamiltonian is $H = p^2/2m + \varphi(x)$, one finds

$$-\frac{(1-q)\beta}{4m^3}\frac{d\beta}{dx}p^5 + \frac{(1-q)(3-q)}{4m^3}\zeta\beta^2 p^4 + \frac{1}{2m^2}\frac{d\beta}{dx}[1 - 2(1-q)\beta\varphi]p^3$$

$$+ \frac{\beta}{m^2}\left[\frac{1+q}{2}B\beta - (2-q)\zeta(1-(1-q)\beta\varphi)\right]p^2 + \frac{\varphi}{m}\frac{d\beta}{dx}[1-(1-q)\beta\varphi]p$$

$$- [1-(1-q)\beta\varphi]\frac{B\beta}{m} + [1-(1-q)\beta\varphi]^2\frac{\zeta}{m} = 0. \tag{20}$$

Very clearly, from this equation one can obtain the following three identities to be satisfied for the Tsallis distribution $f \sim [1-(1-q)\beta(p^2/m + \varphi(x))]^{\frac{1}{1-q}}$. Namely, if the Tsallis distribution is a stationary-state solution of the Fokker-Planck equation, then it must fulfill the three identities,

$$(i)\ \frac{d\beta}{dx} = 0, \tag{21}$$

$$(ii)\ \zeta = 0, \tag{22}$$

$$(iii)\ B = 0. \tag{23}$$

The three identities can determine possible dynamics compatible with the Tsallis distribution function. Eqs.(21)-(23) imply the dynamics of an isothermal process, or a thermal equilibrium system, with no friction and no noise. In this case, the corresponding dynamical equation becomes

$$\frac{dp}{dt} = -\frac{d\varphi}{dx} + F_p(t), \qquad (24)$$

and the noise is irrelated, $<F_p(t)F_p(t')>=0$. Eq.(24) is the dynamical evolution equation of the system governed only by the potential field. When we regard the potential function $\varphi$ as the gravitational one, Eq.(24) characterizes the dynamical evolution of the particle in an self-gravitating system, e.g. the dark matter is such a case. There have been quite a lot of works on the applications of nonextensive statistics to self-gravitating systems, to many-body systems, and to others [1-4]. Our results lead to the conclusion that the current form of the Tsallis distribution cannot describe any nonequilibrium dynamics of the Langevin equation (14) of the systems governed by *any* potential, except the simple isothermal situation, viz. the dynamical system governed by Eq.(24) when it reaches to a thermal equilibrium state. So some applications existed using the Tsallis distribution need to be reconsidered. The examples of the Tsallis distribution behaving only as an isothermal distribution in the above dynamics were reported recently in the N-body simulations [16] and also in the theoretical aspects for the self-gravitating systems [11].

Nothing more can be found for the Tsallis distribution from Eq.(20), except the three properties. We note that if there is no noise or friction, the Fokker-Planck equation reduces to the standard Liouville equation for the Hamiltonian, $H = p^2/2m + \varphi(x)$. With noise and friction, the stationary-state solution can be Boltzmann distribution, $f \sim \exp(-\beta H)$. The statistical property for a thermal equilibrium system, $d\beta/dx = 0$, with $B\beta = \zeta$ (the fluctuation-dissipation theorem) in the dynamics of the Langevin equations can be described by Boltzmann distribution in the conventional statistical mechanics. From this point of view, we cannot find that the Tsallis distribution is necessary for the statistical description of the dynamical system governed by the Langevin equations, Eq.(14), with *any* potential function $\varphi(x)$ because one has not specified whether or not the potential is long-range one. In other words, there is no possible non-equilibrium dynamics of the Langevin equations

Eq.(14) that should use the Tsallis distribution in nonextensive statistical mechanics for the statistical description.

In conclusion, we have used the Fokker-Planck equation to study the dynamical property of the Tsallis distribution. We start with a general Langevin equation of the dynamical system with an *arbitrary* potential function, the noise is Gaussian and delta-function correlated, and the noise-free part is Markovian. We assume the Tsallis distribution is a stationary-state solution of the Fokker-Planck equation and then search for if it is a physical solution compatible with the dynamical functions in the Langevin equation. We show that no possible nonequilibrium dynamics of the Langevin system governed by *any* potential can use the Tsallis distribution for the statistical description, among which the dark matter is one example of such cases. The Tsallis distribution only stands for a simple isothermal or thermal equilibrium situation with no friction and no noise.

**Acknowledgement**

This work is supported by the National Natural Science Foundation of China under grant No.10675088.**References**

[1] C. Tsallis, Introduction to Nonextensive Statistical Mechanics --- Approaching a Complex World, Springer, New York, 2009.

[2] C. Tsallis, in: G. Contopoulos, P. A. Patsis (Eds.), Chaos in Astronomy, Astrophys. Space Sci. Proceedings, Springer, Berlin Heidelberg, 2009, P.309-318.

[3] M.Gell-Mann, C.Tsallis, Nonextensive Entropy --- Interdisciplinary Applications, Oxford University Press, New York, 2004.

[4] S. Abe, Y. Okamoto, Nonextensive Statistical Mechanics and its Applications, Springer, Berlin, 2001.

[5] J. P. Boon, C. Tsallis, Nonextensive Statistical Mechanics: New Trends, New Perspective, Europhys. News **36/6**(2005)185-228.

[6] R. Balian, M. Nauenberg, Europhys. News **37**(2006)9; F. Bouchet, T. Dauxois, S.7